\documentclass[%
 reprint,
 amsmath,amssymb,
 aps,twocolumn,prl
]{revtex4-1}

\usepackage{graphicx}
\usepackage{hyperref}

\newcommand{\eqn}[1]{\begin{eqnarray} #1 \end{eqnarray}}
\newcommand{\tit}[1]{\textit{#1}}
\newcommand{\tbf}[1]{\textbf{#1}}

\begin{document}

\title{Comment on ``The notion of locality in relational quantum mechanics"}

\author{Jacques Pienaar}
\affiliation{
 International Institute of Physics, Universidade Federal do Rio Grande do Norte, Campus Universitario, Lagoa Nova, Natal-RN 59078-970, Brazil.
}

\date{\today}



\begin{abstract}
A recent paper [P. Martin-Dussaud, C. Rovelli, F. Zalamea, arXiv:1806.08150] has given a lucid treatment of Bell's notion of local causality within the framework of the relational interpretation of quantum mechanics. However, the authors went on to conclude that the quantum violation of Bell's notion of local causality is no more surprising than a common cause. Here, I argue that this conclusion is unwarranted by the authors' own analysis. On the contrary, within the framework outlined by the authors, I argue that the implications of the relational interpretation are much more radical. 
\end{abstract}

\maketitle


\section{BACKGROUND}

In \cite{MARTY}, an analysis of Bell's notion of \tit{local causality} was given from the standpoint of the relational interpretation of quantum mechanics \cite{RELATIONAL}. The authors made the following points, which I take to be convincing. To paraphrase:\\

\noindent (1) Bell's notion of local causality is a statement about the \tit{beables} of the theory (i.e. the candidate elements of reality);\\
(2) The definition and properties of the beables, and hence the applicability of local causality, depends on one's interpretation of quantum mechanics;\\
(3) According to the relational interpretation, all physically meaningful beables relative to an observer are located within that observer's past light-cone;\\
(4) In particular, Bell's criterion of local causality can only be meaningfully defined (and violated) relative to an observer whose past light cone encompasses all relevant beables.\\

An \tit{observer} in this context refers to a physical system that is localized to a time-like trajectory of finite or infinite extent in space-time (a world-line segment) \footnote{Note that there may be additional requirements for a physical system to qualify as an \tit{observer}, such as being macroscopic, having sense-organs, etc; sufficient conditions for observerhood are left unspecified here.}. It will be useful to restrict our attention to \tit{finite observers}, whose world-lines stretch from an initial `starting event' to a final `terminal event'. The `past light cone' referenced in point (3) above is then understood to mean the past light-cone of the observer's terminal event, which encompasses the past light-cones of all other events on the observer's world-line (This elaboration was not made in \cite{MARTY} but I take it to be consistent with their analysis).

\begin{figure}[!htb]
\centering\includegraphics[width=0.8\linewidth]{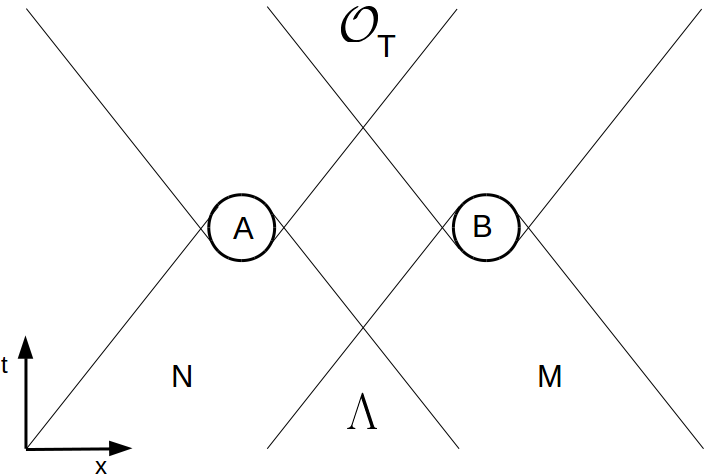}
\caption{A space-time diagram showing the approximate location of the beables $\{ A,B,N,M,\Lambda \}_{ \mathcal{O} }$ relative to an observer $\mathcal{O}$, according to \cite{MARTY}. The circles are the space-time regions where local measurements take place, and the thin lines delineate their past and future light-cones. Since $\mathcal{O}$ considers $\{ A,N \}_{ \mathcal{O} }$ to be space-like separated from $\{ B,M \}_{ \mathcal{O} }$, a violation of Bell's inequalities for these beables implies a violation of local causality relative to this observer.}
\label{fig:beaBells}
\end{figure}

We now review the last point (4), and its analysis in \cite{MARTY}. A Bell measurement scenario is defined by a set of beables $\{ A,B,N,M,\Lambda \}_{ \mathcal{O} }$. The beables $A,B$ refer to the outcomes of two measurements on different parts of a quantum system. $N$ is a set of beables that are in the causal past of $A$ (but not $B$), which includes the measurement setting relevant to $A$. Similarly $M$ is a set of beables that are in the causal past of $B$ (but not $A$), which includes the measurement setting relevant to $B$. $\Lambda$ represents a set of beables in the common causal past of $A,B$, including the preparation of the quantum system. All of these beables are interpreted as being `physically meaningful' relative to the observer $\mathcal{O}$ and are therefore all located in this observer's past light-cone. It remains to be specified exactly \tit{where} in the past light-cone they are situated, which will be the main point of contention taken up later on; for the time being let us follow \cite{MARTY} by locating the beables as shown in Fig. \ref{fig:beaBells}. 

The authors of \cite{MARTY} do not specify the entire world-line of $\mathcal{O}$ because their analysis only depends on the observer's `past light cone' and hence only on the terminal event of the world-line, labelled $\mathcal{O}_T$ in the diagram. For any observer terminating at this event, the beables $A,B,N,M,\Lambda$ will be in their past light cone and thus potentially physically meaningful. Any such observer can potentially witness the experimental violation of Bell's inequalities and hence reject Bell's condition of local causality. The latter condition may be expressed relative to the observer $\mathcal{O}$ as the following condition on the observed probabilities:
\eqn{
P(a | \lambda, n, b )_{\mathcal{O}} = P(a | \lambda, n )_{\mathcal{O}} \, ,
}
where the lowercase letters represent the specific values of the associated beables. 
It is important to notice that the relativisation of the beables to the observer $\mathcal{O}$ has apparently gotten us no further towards explaining or avoiding this violation of local causality -- it has only made it a problem for a smaller class of observers than is usually considered, namely those terminating at an event in the future light-cones of both measurements. For these observers, however, the problem seems just as strong as it would be without the relational interpretation.

At this juncture, the authors of \cite{MARTY} choose to bite the bullet and give up on local causality. The faulty assumption, they argue, is the idea that the correlations among the effects of a common cause should disappear when the common cause is conditioned upon. This assumption, also known variously as `Reichenbach's \tit{quantitative} principle of common causes' \cite{CAVLAL}, `factorization' and `decorrelating explanation' \cite{WISECAV}, is often argued to be based upon the assumption that the underlying physics is deterministic. Authors sympathetic to the idea that quantum mechanics is a fundamentally non-deterministic theory have therefore found good reason to reject this assumption, and thus to reject Bell's local causality as being a meaningful notion of `locality' for quantum mechanics. 

This approach, while perfectly respectable, is quite independent of the relational interpretation. Indeed, if one does not regard local causality as being the proper notion of locality for indeterministic beables (as advocated in \cite{MARTY}), then it hardly seems to matter whether these beables are relational or not: all of the heavy lifting has already been done by the appeal to `indeterminism', and there is nothing that conceptually requires relationalism.

I would like to suggest that the authors of \cite{MARTY} have given up too easily. In the remainder of this comment, I show that a more careful examination of beables in the relational interpretation allows us to avoid violations of Bell's notion of local causality altogether. There is no need to reject the factorization principle of common causes -- the relational interpretation can dissolve the problem without assistance. This solution comes at a price: one must be willing to take the relational interpretation of beables seriously and pursue it to its logical limit. As we will see, this has some radical consequences that are likely to make even the authors of \cite{MARTY} uncomfortable.

\section{When does a beable begin to be?}

Imagine that you, a keen astronomer, are the observer $\mathcal{O}$. You are the witness to a shocking event, let us say a murder, that occurs in front of your eyes as you are watching events on the Mars base through your telescope. The perpetrator and victim are both known to you: the troubled relationship between Alice and Bob is no secret among the close-knit quantum information community. But now Alice has taken things past the point of no return, plunging a knife deep into Bob's heart. Let us call the violent act $M$, and let $W$ be the event that light from the murder lands upon your startled eyes. The terminal event $\mathcal{O}_T$ of your observer-world-line occurs shortly after $W$ (after which we may suppose you are a wholly different person). Thus $M$ clearly satisfies the criterion of being within your past light-cone. The question is, where in your past light-cone did $M$ actually happen? 

Hurriedly, you tear a map of Minkowski space-time from the wall, intending to mark with your pencil the exact space-time event of the murder $M$. Logic compels you to say that it happened on Mars, and so you trace backwards the path of the photons from the event $W$ -- whose co-ordinates are obviously known to you -- all the way back to their origin on Mars' surface, to a location just outside a popular Mars disco.

Then doubt sets in -- on what grounds can you say that $M$ really happened? The only fact directly accessible to you is the fact that photons carrying the image of a murder impacted your retinas at event $W$. Surely you are entitled to reason that this was caused by an actual murder that took place elsewhere, but this only seems to support the notion that $M$ is not itself a beable, but rather it is something more of an analytic fact deduced from the actual beables. These latter would include $W$, plus any relevant memories from your life history, but not anything resembling a direct experience of (i.e. physical interaction with) the actual murder $M$. For you, there is really no $M$ in the sense of `beable', but only $W$, which occurs of necessity right where you were standing at the telescope, and not 64 million kilometers away on Mars.

Well, what about Bob? $M$ is surely a beable to him. Moreover, on being murdered, Bob might reasonably expect light from this event to reach your eyes informing you of the betrayal -- he might even be \tit{certain} that you are in the habit of observing Mars at precisely the moment that light from the foul deed would reach Earth. But to him, it is $M$ that is the real beable and $W$ that is only an abstract rational deduction from it. We are only taking seriously what the relational interpretation says: beables arise in local interactions between systems. $W$ arises from the interaction between the light arriving at your telescope and you, and so it is physically meaningful to you (but not to Bob). $M$ arises from the deadly interaction between Alice and Bob, and is physically meaningful to them, but not to you.

Let me pause to examine a possible objection to this line of thought. Surely, there is an objective sense in which $M$ \tit{actually happened} that goes beyond the matter of whether light from $M$ reached $\mathcal{O}$. Suppose that at the critical moment the International Space Station happened to pass between you and Mars, obstructing the view through your telescope, and preventing any light rays from Mars from reaching you. Nevertheless, Bob is dead, and will be dead when you next meet him (in the morgue). Let us therefore suppose that there is nothing that happens within the past light-cone of an observer that is \tit{not} a beable for that observer, and strengthen statement (3) to the following:\\

(3): Light-cone version: All physically meaningful beables relative to an observer are located within that observer's past light-cone, \tit{and all beables in the observer's past light cone are physically meaningful to that observer}.\\

This objection allows us to maintain some measure of observer relativism but without going the whole hog. It allows us to say that a beable comes into existence for an observer merely when it is possible `in principle' for information about it to reach the observer. 

To reply to this, it is enough to probe into the deeply unsatisfactory concessions lurking within that phrase \tit{`in principle'}. The past light-cone refers to light travelling in a vacuum, yet nowhere in the universe is a true vacuum to be found. Thus, the rule (3)' would have us admit that beables become physically meaningful to us a moment before light actually reaches us in any realistic circumstance, having been slowed down by an intervening medium. They are said to be physically meaningful only because the light \tit{would} have reached us at the same time \tit{if only the intervening matter hadn't been there}.

If the absurdity of that counterfactual isn't troubling enough, consider that this approach denies any possible connection between beables and local physical interactions. Suppose I were to seal up an observer in a tank that maintained the quantum state of their body at the highest possible purity, preventing nearly all their physical interactions with the external environment. Relative to that observer, is it really sensible to insist on the physical meaningfulness of external events that cannot possibly affect them in any way? Note that one cannot escape from this bind by saying that these events would be relevant to the observer's experience in a possible future in which they are released from the tank. To talk meaningfully about an observer, we must designate a terminal event $\mathcal{O}_T$ after which the observer effectively ceases to exist, for it is this event that defines the past light-cone as a strict subset of space-time. If the observer is kept in a tank until this point, the argument carries through unchanged. One could try appealing to a counterfactual -- that the beables are physically meaningful by virtue of the consequences they \tit{would} have if the observer \tit{weren't} in the tank -- but this leads us astray from the original relational interpretation, which assigns beables to physical interactions that \tit{actually} occur, not to hypothesized local interactions that \tit{might} occur.

There seems, in short, no way to avoid the conclusion that the space-time location of beables for an observer $\mathcal{O}$ occur at the space-time location of the relevant physical interaction with that observer, which means \tit{along the observer's world-line}. Thus we should rather strengthen (3) to the following:\\

(3): World-line version: All physically meaningful beables relative to an observer are located along that observer's past \tit{world-line}.\\

This rather simple manoeuvre completely transforms the nature of quantum theory's conflict with local causality. In effect, there is trivially no conflict, because for any real observer $\mathcal{O}$, the beables that correspond to that observer's experience of the two measurement events cannot be space-like separated. Thus, in cases where correlations do not admit factorization by a purported common cause, the criterion of local causality can present no objection to an explanation in terms of a causal effect from one to the other. Let us spell this out in a little more detail. 

\begin{figure}[!htb]
\centering\includegraphics[width=0.8\linewidth]{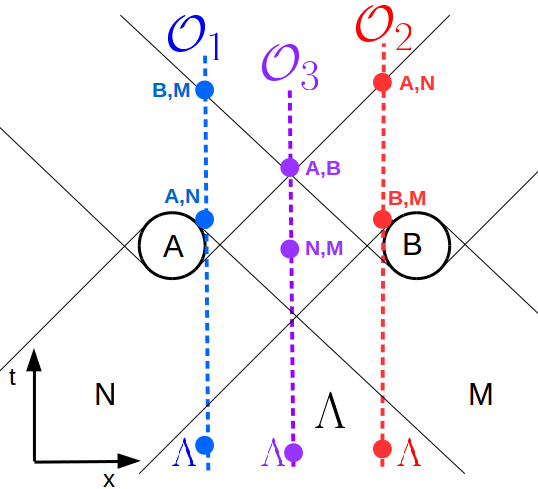}
\caption{The space-time locations of the beables relative to three different observers. For each observer, the physically meaningful beables are located on that observer's world line (dashed lines). The relative beables are represented by solid circles, colour-coded to match their corresponding observer. The black labels $A,B,N,M,\Lambda$ are not beables relative to any individual observer, but represent \tit{abstract} beables constructed from the combined information of all three observers.}
\label{fig:worldlineBell}
\end{figure}

In general, the observers for whom the key beables $A,B,N,M,\Lambda$ are physically meaningful can be classified into three categories (see Fig. \ref{fig:worldlineBell}). In the first category are those such as the observer $\mathcal{O}_1$ in the figure, for whom the beables $\{ A,B,N,M,\Lambda \}_{\mathcal{O}_1}$ occur in a time-like sequence in which $\{ A,N \}_{\mathcal{O}_1}$ come before $\{ B,M \}_{\mathcal{O}_1}$. There may yet be reasons to doubt that correlations between them are due to an actual \tit{causal influence} from the former to the latter (more on this in a minute), but such a causal explanation can no longer be excluded on the grounds of violating Bell's notion of local causality, since for the observer in question the relevant beables are time-like (rather than space-like) separated events. A symmetric argument can be made for observers like $\mathcal{O}_2$ for whom $\{ B,M \}_{\mathcal{O}_2}$ are in the time-like past of $\{ A,N \}_{\mathcal{O}_2}$. The third class of observers are those like $\mathcal{O}_3$, for whom the beables $\{ N,M \}_{\mathcal{O}_3}$ first become physically meaningful at a single space-time event, and subsequently $\{ A,B \}_{\mathcal{O}_3}$ at a later event along the world-line. Again, $\mathcal{O}_3$ is free to suppose a causal explanation (in either direction) between $\{ A,N \}_{\mathcal{O}_3}$ and $\{ B,M \}_{\mathcal{O}_3}$ to help explain the observed violations of Bell inequalities, since the beables in question are only time-like separated.

Here a critic can raise a serious objection, for if, say,  $\mathcal{O}_1$ wishes to explain the violation of Bell inequalities by appealing to a causal influence from $N_{\mathcal{O}_1}$ to $B_{\mathcal{O}_1}$, then she is also obliged to explain why this supposed influence does not permit her to transmit a signal to $B_{\mathcal{O}_1}$ by manipulating $N_{\mathcal{O}_1}$, as it is well-known that quantum mechanics forbids such signalling. Fortunately, it is precisely here that the relational interpretation comes through with a beautiful explanation, which only requires an expansion of the domain of the \tit{relativity principle} to the following:\\

\noindent \tbf{Extended relativity principle:} In cases where different observers dispute the direction of causality, the laws of physics do not permit any experiment that could favour the view of one observer over another.\\

If it were possible for $\mathcal{O}_1$ to signal to $B_{\mathcal{O}_1}$ by manipulating $N_{\mathcal{O}_1}$, this fact would force $\mathcal{O}_2$ to agree that $N$ is in the time-like past of $B$, which agrees better with $\mathcal{O}_1$'s placement of the beables than that of $\mathcal{O}_2$. This would seem to favour $\{ A,B,N,M,\Lambda \}_{\mathcal{O}_1}$ as being the `true beables' instead of $\{ A,B,N,M,\Lambda \}_{\mathcal{O}_2}$, which would violate the extended relativity principle. Hence, if the latter is to be respected, signalling cannot be possible.

The extended relativity principle admittedly stretches the imagination, as we are not used to thinking of causal relations as being observer-dependent in the same way that simultaneity is, or the time-ordering of space-like separated events. Yet some recent work has given us reasons to think of causality as a relative concept \cite{GUERIN, JACQUES, ORESHKOV, PRICE}. The extended relativity principle would have us distinguish two types of causes in nature: those whose existence and direction is agreed upon by all observers (and hence which can be used for signalling) and those whose existence is agreed upon by all observers but whose \tit{direction} is observer-relative, and these kinds of causal influence cannot be used for signalling. Bell-inequality violation can therefore be interpreted as the discovery of this second category of causal relations in nature.

\section{Rescuing objectivity}

The world-line version of (3) seems to force on us a picture of reality that is hopelessly fragmented: if each observer has their own personal beables, then what is it that unites their experiences and gives rise to the emergence of an objective reality? True enough, ten people looking at the same tree will all see something different, but the `world-line version' of (3) seems to commit us to saying that the tree is an illusion, and there are in fact ten trees, one for each observer. Something is clearly missing from this account -- we can't seem to see the tree for the forest.

The solution is to recognize that there are two aspects to `reality', namely a personal aspect and a collective aspect. The personal reality of an observer is just what we have been describing, namely, the beables that are strung out along that observer's own world-line, which represent the private experiences of the observer. However, our reality is constructed not only from our own observations, but also those made by other people. In the example of the Bell experiment, consider a meeting between the observers $\mathcal{O}_1$ and $\mathcal{O}_2$. These observers may find it compelling to make an identification between the beable $\{ A\}_{\mathcal{O}_1}$ (seen by $\mathcal{O}_1$) and the beable $\{ A\}_{\mathcal{O}_2}$ (seen by $\mathcal{O}_2$), by choosing to regard these as different points of view of one and the same \tit{abstract beable} $A$. This $A$ really is an abstraction, because it is derived from a pair of beables that are not jointly `physically meaningful' for any single observer, at least not in the same strong sense as each individual's private beables. Yet it does have a claim to being physically meaningful for the \tit{collective} of observers whose disparate beables it unites. This fusion of several disparate `realities' into a single `collective reality' is only possible if one makes an identification between the beables of different observers, by imagining them to represent different points of view of the same `thing'. The `thing' in question (the \tit{abstract beable}) is only as real as much as the proposed identification of beables between observers makes sense.

If we grant that such identifications can be made (on whatever grounds), we can then imagine filling space-time with observers and allowing them to communicate with each other, pooling their separate experiences to reconstruct their past light-cone using abstract beables whose space-time location is the same for all members of the community. Thus, despite initial appearances, the world-line version of (3) does not commit us to a hopeless subjectivity, but leaves enough room for observers to reach agreement on the existence of `objective', perhaps better named \tit{abstract} or `inter-subjective' beables.

\section{Some holes in the plot}

I have argued that the relational interpretation of beables in \cite{MARTY} does not necessarily run into conflict with Bell's notion of local causality, because for any actual observer in space-time, the relevant beables corresponding to each measurement only become physically meaningful after the observer interacts with some system carrying information about them, such as a light beam, and thus are always time-like (and not space-like) separated for that observer. However, the violation of Bell inequalities between these time-like separated beables still conflicts with a common cause explanation, if we adhere to principle of common-cause factorization. Each observer is then forced to posit a direct causal relationship between the time-like separated beables, and as we have seen, they may not agree on the direction of this cause. But this also raises a new question: what is the physical nature of this causal connection? Can we make it explicit within a dynamical model? For instance, can we define a state of the system relative to $\mathcal{O}_1$ such that manipulating $\{ N \}_{\mathcal{O}_1}$ can be seen to dynamically affect the relative state of $\{ B,M \}_{\mathcal{O}_1}$? Such an account, if possible, would lend more credibility to this idea.

It is also important to emphasize that the entire argument depends upon the assumption that Bell's notion of \tit{local causality} applies only to beables as I have defined them (as direct physical interactions with an observer) and not to what I have been calling \tit{abstract beables}. This move can easily be attacked. Consider that, though they may be \tit{abstract}, the latter are no less \tit{beables} insofar as they are the very building blocks of what the observers call `objective reality' as explained in the preceding section, and thus may well be the appropriate subject matter of \tit{local causality}. As an analogy, consider the case of the classical electromagnetic field. The beables are represented by the local measurements of this field performed by various observers in various places, but what of the values of the field in parts of space where it was not measured by any observer? We nevertheless infer that the field has a value there and everywhere else in space, and by this inference we introduce the \tit{electromagnetic field} as an \tit{abstract beable} (or abstract-beable-valued function if you prefer), which constitutes an element of our shared reality. Now when we discuss \tit{local causality} in the context of electromagnetism, we can ask whether this constraint applies only to the isolated and patchy parts of the field that were actually measured (the field beables), or whether it applies to the electromagnetic field as a field of abstract beables filling all of space. Surely, it is the latter. If we continue to extend \tit{local causality} to abstract beables in a like manner when considering quantum systems, then our argument given above cannot save \tit{local causality} from Bell inequality violations. In this case, we must admit that the relational interpretation has made little headway in solving this problem, and a move such as rejecting common-cause factorization as in Ref. \cite{MARTY} may be warranted. 

Finally, a gap remains regarding the precise way in which observers are supposed to construct a coherent picture of reality. Since we have given up on any external reality as the origin of our sense-experiences, then why should we expect even the appearance of such a reality to emerge between us? It may seem that we are assuming an unexplained miracle by supposing that the phenomena of our private experiences should admit identifications to be made between us (why should the tree that I see look anything similar to the one that you see?) Without making this assumption, the notion of an objective world and indeed science itself would be impossible. True enough, but note that the realist faces a similar problem, since he must take it on faith that an orderly external world exists in the first place -- after all, why should it? The mystery was there all along, we have only relocated it by emphasizing that it is not a question about how things came to exist prior to, and independent of, observation, but rather a question of how things come to exist \tit{through the physical process of observation}.

By thus refocusing the problem, we may have gotten closer to resolving it. Recent related work by M. M\"{u}ller \cite{MUELLER} has argued that the emergence of an apparent reality for all observers can derived from the principle that Solomonoff induction correctly predicts each observer's future observations. Thus, there may be an answer to the question of why a single objective world seems to exist, if we only let go of the idea that it must exist \tit{a priori}. Then objective reality is rather more like a tapestry collectively woven by many observers; each of us contributes one patch of it made of our own private experiences, and these are stitched together by our physical interactions with one another, as dictated by the rules of quantum mechanics. Perhaps this is the real lesson to be drawn from the relational interpretation. \\

\acknowledgments
I thank G. Barreto Lemos for helpful feedback on an earlier draft.

\end{document}